\documentclass[conference]{IEEEtran}

\usepackage{cite}

\usepackage{graphicx}
\graphicspath{{/home/aawan/Documents/Ahsan_Research_Papers/PhD_papers/HPDC_extended_arvix_submission/}}
\DeclareGraphicsExtensions{.jpg}

\usepackage{array}
\usepackage{amssymb}
\usepackage{amsmath}

\usepackage{caption}
\ifCLASSOPTIONcompsoc
\usepackage[caption=false,font=normalsize,labelfont=sf,textfont=sf]{subfig}
\else
\usepackage[caption=false,font=footnotesize]{subfig}
\fi

\usepackage{multirow}
\usepackage[export]{adjustbox}

\PassOptionsToPackage{hyphens}{url}\usepackage{hyperref}

\usepackage{comment}

\begin{document}

\title{Architectural Impact on Performance of In-memory Data Analytics: Apache Spark Case Study}

\author{

\IEEEauthorblockN{
Ahsan Javed Awan\IEEEauthorrefmark{1},  
Mats Brorsson\IEEEauthorrefmark{1}, 
Vladimir Vlassov\IEEEauthorrefmark{1} and 
Eduard Ayguade\IEEEauthorrefmark{2}
}

\IEEEauthorblockA{\IEEEauthorrefmark{1}KTH Royal Institute of Technology,\\
Software and Computer Systems Department(SCS),\\
\{ajawan, matsbror, vladv\}@kth.se}

\IEEEauthorblockA{\IEEEauthorrefmark{2}Technical University of Catalunya (UPC),\\
Computer Architecture Department,\\
eduard@ac.upc.edu}


}


\maketitle
\begin{abstract}
While cluster computing frameworks are continuously evolving to provide real-time data analysis capabilities, Apache Spark has managed to be at the forefront of big data analytics for being a unified framework for both, batch and stream data processing. However, recent studies on micro-architectural characterization of in-memory data analytics are limited to only batch processing workloads. We compare micro-architectural performance of batch processing and stream processing workloads in Apache Spark using hardware performance counters on a dual socket server. In our evaluation experiments, we have found that  batch processing are stream processing workloads have similar micro-architectural characteristics are bounded by the latency of frequent data access to DRAM. For data accesses we have found that simultaneous multi-threading is effective in hiding the data latencies. We have also observed that (i) data locality on NUMA nodes can improve the performance by 10\% on average and(ii) disabling next-line L1-D prefetchers can reduce the execution time by up-to 14\% and (iii) multiple small executors can provide up-to 36\% speedup over single large executor

\end{abstract}
\IEEEpeerreviewmaketitle

\section{Introduction}
With a deluge in the volume and variety of data collecting, web enterprises (such as Yahoo, Facebook, and Google) run big data analytics applications using clusters of commodity servers. 
However, it has been recently reported that using clusters is a case of over-provisioning since a majority of analytics jobs do not process really data sets and that modern scale-up servers are adequate to run analytics jobs~\cite{Scale_up_vs_Scale_out_for_Hadoop}. Additionally, commonly used predictive analytics such as  machine learning algorithms, work on filtered datasets that easily fit into memory of modern scale-up servers. Moreover the today's scale-up servers can have CPU, memory and persistent storage resources in abundance at affordable prices. Thus we envision small cluster of scale-up servers to be the preferable choice of enterprises in near future. 

While Phoenix~\cite{Phoenix_Rebirth}, Ostrich~\cite{Tiled_mr} and Polymer~\cite{Polymer} are specifically designed to exploit the potential of a single scale-up server, they do not scale-out to multiple scale-up servers. Apache Spark~\cite{Spark} is getting popular in the industry because it enables in-memory processing, scales out to large number of commodity machines and provides a unified framework for batch and stream processing of big data workloads. However its performance on modern scale-up servers is not fully  understood. Recent studies ~\cite{performance_spark, performance_spark_volume} characterize the performance of in-memory data analytics with Spark on a scale-up server but they are limited in two ways. Firstly, they are limited only to batch processing workloads and secondly, they do not quantify the impact of NUMA, Hyper-Threading and hardware prefetchers on the performance of Spark workloads. Knowing the limitations of modern scale-up servers for in-memory data analytics with Spark will help in achieving the future goal of improving the performance of in-memory data analytics with Spark on small clusters of scale-up servers. 

Our contributions are:

\begin{itemize}
\item We characterize the micro-architectural performance of Spak-core, Spark Mllib, Spark SQL, GraphX and Spark Streaming.
\item We quantify the impact of data velocity on micro-architectural performance of Spark Streaming.
\item We analyze the impact of data locality on NUMA nodes for Spark. 
\item We analyze the effectiveness of Hyper-threading and existing prefetchers in Ivy Bridge server to hide data access latencies for in-memory data analytics with Spark. 
\item We quantify the potential for high bandwidth memories to improve the the performance of in-memory data analytics with Spark.  
\item We make recommendations on the configuration of Ivy Bridge server and Spark to improve the performance of in-memory data analytics with Spark

\end{itemize}

The rest of this  paper is organised as follows. Firstly, we provide background and formulate hypothesis in section 2. Secondly, we discuss the experimental setup in section 3, examine the results in section 4 and discuss the related work in section 5. Finally we summarize the findings and give recommendations the in section 6.  

\section{Background}
\subsection{Spark}
Spark is a cluster computing framework that uses Resilient Distributed Datasets (RDDs)~\cite{Spark} which are immutable collections of objects spread across a cluster. Spark programming model is based on higher-order functions that execute user-defined functions in parallel. These higher-order functions are of two types: ``Transformations'' and ``Actions''. Transformations are lazy operators that create new RDDs, whereas Actions launch a computation on RDDs and generate an output. When a user runs an action on an RDD, Spark first builds a DAG of stages from the RDD lineage graph. Next, it splits the DAG into stages that contain pipelined transformations with narrow dependencies. Further, it divides each stage into tasks, where a task is a combination of data and computation. Tasks are assigned to executor pool of threads. Spark executes all tasks within a stage before moving on to the next stage. Finally, once all jobs are completed, the results are saved to file systems.

\subsection{Spark MLlib}
Spark MLlib is a machine learning library on top of Spark-core.  It contains commonly used algorithms related to collaborative filtering, clustering, regression, classification and dimensionality reduction.

\subsection{Graph X}
GraphX enables graph-parallel computation in Spark. It includes a collection of graph algorithms. It introduces a new Graph abstraction: a directed multi-graph with properties attached to each vertex and edge. It also exposes a set of fundamental operators (e.g., aggregateMessages, joinVertices, and subgraph) and optimized variant of the Pregel API to support graph computation.

\subsection{Spark SQL}
Spark SQL is a Spark module for structured data processing. It provides Spark with additional information about the structure of both the data and the computation being performed. This extra information is used to perform extra optimizations. It also provides  SQL API, the DataFrames API and the Datasets API. When computing a result the same execution engine is used, independent of which API/language is used to express the computation.  

\subsection{Spark Streaming}
Spark Streaming~\cite{DStreams} is an extension of the core Spark API for the processing of data streams. It provides a high-level abstraction called discretized stream or DStream, which represents a continuous stream of data. Internally, a DStream is represented as a sequence of RDDs. Spark streaming can receive input data streams from sources such like Kafka, Twitter, or TCP sockets. It then divides the data into batches, which are then processed by the Spark engine to generate the final stream of results in batches. Finally, the results can be pushed out to file systems, databases or live dashboards.  

\subsection{Garbage Collection}
Spark runs as a Java process on a Java Virtual Machine(JVM). The JVM has a heap space which is divided into young and old generations.  
The young generation keeps short-lived objects while the old generation holds objects with longer lifetimes. The young generation is further divided into eden, survivor1 and survivor2 spaces. When the eden space is full, a minor garbage collection (GC) is run on the eden space and objects that are alive from eden and survivor1 are copied to survivor2. The survivor regions are then swapped. If an object is old enough or survivor2 is full, it is moved to the old space. Finally when the old space is close to full, a full GC operation is invoked.

\subsection{Spark on Modern Scale-up Servers}

Dual socket servers  are becoming commodity in the data centers, e.g Google warehouse scale computers have been optimized for NUMA machines~\cite{google_numa}. Moder scale-up servers like Ivy Bridge machines are also part of Google data centers as recent study~\cite{profiling_warehouse} presents a detailed micro-architectural analysis of Google data center jobs running on the Ivy Bridge servers. Realizing the need of the hour, the inventors of Spark are also developing cache friendly data structures and algorithms under the project name Tungsten ~\cite{tungsten} to improve the memory and CPU performance of Spark applications on modern servers

Our recent efforts on identifying the bottlenecks in Spark~\cite{performance_spark, performance_spark_volume} on Ivy Bridge machine shows that (i) Spark workloads exhibit poor multi-core scalability due to thread level load imbalance and work-time inflation, which is caused by frequent data accesses to DRAM and (ii) the  performance of Spark workloads deteriorates severely as we enlarge the input data size due to significant garbage collection overhead. However, the scope of work is limited to batch processing workloads only, assuming that Spark streaming would have same micro-architectural bottlenecks. We revisit this assumption in this paper. 

Simulatenous multi-threading and hardware prefectching are effective ways to hide data access latencies and additional latency over-head due to accesses to remote memory can be removed by co-locating the computations with data they access on the same socket. 

One reason for severe impact of garbage collection is that full generation garbage collections are triggered frequently at large volumes of input data and the size of JVM is directly related to Full GC time. Multiple smaller JVMs could be better than a single large JVM. 

In this paper, we answer the following questions concerning in-memory data analytics running on modern scale-up servers using the Apache Spark as a case study. Apache Spark defines the state of the art in big data analytics platforms exploiting data-flow and in-memory computing.

\begin{itemize}
\item Does micro-architectural performance remain consistent across batch and stream processing data analytics?
\item How does data velocity affect micro-architectural performance of in-memory data analytics with Spark?
\item How much performance gain is achievable by co-locating the data and computations on NUMA nodes for in-memory data analytics with Spark?  
\item Is simultaneous multi-threading effective for in-memory data analytics with Spark? 
\item Are existing hardware prefetchers in modern scale-up servers effective for in-memory data analytics with Spark?
\item Does in-memory data analytics with Spark experience loaded latencies (happens if bandwidth consumption is more than 80\% of sustained bandwidth)
\item Are multiple small executors (which are java processes in Spark that run computations and store data for the application) better than single large executor?
\end{itemize}

\section{Methodology}
Our study of architectural impact on in-memory data analytics is based on  an imperial study of performance of batch and stream processing with Spark using representative benchmark workloads. We have performed several series of experiments, in which we have evaluated impact of each of the architectural features, such as  data locality in NUMA, HW prefetchers, and hyper-threading, on in-memory data analytics with Spark.

\subsection{Workloads}

This study uses batch processing  and stream processing workloads, described in Table~\ref{batch_workloads} and Table~\ref{stream_workloads} respectively. Benchmarking big data analytics is an open research area, we however chose the workloads carefully. Batch processing workloads are a subset BigdataBench~\cite{BigDataBench} and HiBench~\cite{HiBench} which are highly referenced benchmark suites in Big data domain. Stream processing workloads used in the paper are super set of StreamBench~\cite{streambench} and also cover the solution patterns for real-time streaming analytics~\cite{perera2015solution}.

The source codes for Word Count, Grep, Sort and NaiveBayes are taken from BigDataBench~\cite{BigDataBench}, whereas the source codes for K-Means, Gaussian and Sparse NaiveBayes are taken from Spark MLlib (which is Spark's scalable machine learning library~\cite{mllib}) examples available along with Spark distribution. Likewise the source codes for stream processing workloads are also available from Spark Streaming examples. Big Data Generator Suite (BDGS), an open source tool was used to generate synthetic data sets based on raw data sets~\cite{BDGS}.

\begin{table}[]
\renewcommand{\arraystretch}{1.3}
\centering
\caption{Batch Processing Workloads}
\label{batch_workloads}
\resizebox{\columnwidth}{!}{
\begin{tabular}{l|l|l|l}
\hline
\multicolumn{1}{c|}{\textbf{\begin{tabular}[c]{@{}c@{}}Spark \\ Library\end{tabular}}} & \multicolumn{1}{c|}{\textbf{Workload}} & \multicolumn{1}{c|}{\textbf{Description}} & \multicolumn{1}{c}{\textbf{\begin{tabular}[c]{@{}c@{}}Input \\ data-sets\end{tabular}}} \\ \hline
\multirow{4}{*}{Spark Core} & \begin{tabular}[c]{@{}l@{}}Word Count \\ (Wc)\end{tabular} & counts the number of occurrence of each word in a text file & \multirow{2}{*}{\begin{tabular}[c]{@{}l@{}}Wikipedia \\ Entries\\ (Structured)\end{tabular}} \\ \cline{2-3}
 & Grep (Gp) & \begin{tabular}[c]{@{}l@{}}searches for the keyword The in a text file and filters out the\\ lines with matching strings to the output file\end{tabular} &  \\ \cline{2-4} 
 & Sort (So) & ranks records by their key & \begin{tabular}[c]{@{}l@{}}Numerical\\ Records\end{tabular} \\ \cline{2-4} 
 & \begin{tabular}[c]{@{}l@{}}NaiveBayes \\ (Nb)\end{tabular} & runs sentiment classification & \begin{tabular}[c]{@{}l@{}}Amazon Movie \\ Reviews\end{tabular} \\ \hline
\multirow{5}{*}{Spark Mllib} & \begin{tabular}[c]{@{}l@{}}K-Means \\ (Km)\end{tabular} & \begin{tabular}[c]{@{}l@{}}uses  K-Means clustering algorithm from  Spark Mllib. \\ The benchmark is run for 4 iterations with 8 desired clusters\end{tabular} & \multirow{5}{*}{\begin{tabular}[c]{@{}l@{}}Numerical \\ Records\\ (Structured)\end{tabular}} \\ \cline{2-3}
 & \begin{tabular}[c]{@{}l@{}}Gaussian \\ (Gu)\end{tabular} & \begin{tabular}[c]{@{}l@{}}uses Gaussian clustering algorithm from Spark Mllib. \\ The benchmark is run for 10 iterations with 2 desired clusters\end{tabular} &  \\ \cline{2-3}
 & \begin{tabular}[c]{@{}l@{}}Sparse \\ NaiveBayes\\ (SNb)\end{tabular} & uses NaiveBayes classification alogrithm from Spark Mllib &  \\ \cline{2-3}
 & \begin{tabular}[c]{@{}l@{}}Support Vector\\ Machines (Svm)\end{tabular} & uses SVM classification alogrithm from Spark Mllib &  \\ \cline{2-3}
 & \begin{tabular}[c]{@{}l@{}}Logistic \\ Regression(Logr)\end{tabular} & uses Logistic Regression alogrithm from Spark Mllib &  \\ \hline
\multirow{3}{*}{Graph X} & Page Rank (Pr) & \begin{tabular}[c]{@{}l@{}}measures the importance of each vertex in a graph.  \\ The benchmark is run for 20 iterations\end{tabular} & \multirow{3}{*}{\begin{tabular}[c]{@{}l@{}}Live \\ Journal\\  Graph\end{tabular}} \\ \cline{2-3}
 & \begin{tabular}[c]{@{}l@{}}Connected \\ Components (Cc)\end{tabular} & \begin{tabular}[c]{@{}l@{}}labels each connected component of the graph with the \\ ID of its lowest-numbered vertex\end{tabular} &  \\ \cline{2-3}
 & Triangles (Tr) & \begin{tabular}[c]{@{}l@{}}determines the number of triangles passing through \\ each vertex\end{tabular} &  \\ \hline
\multirow{2}{*}{\begin{tabular}[c]{@{}l@{}}Spark\\ SQL\end{tabular}} & \begin{tabular}[c]{@{}l@{}}Aggregation \\ (SqlAg)\end{tabular} & \begin{tabular}[c]{@{}l@{}}implements aggregation query from BigdataBench \\ using DataFrame API\end{tabular} & \multirow{2}{*}{Tables} \\ \cline{2-3}
 & Join (SqlJo) & \begin{tabular}[c]{@{}l@{}}implements aggregation query from BigdataBench \\ using DataFrame API\end{tabular} &  \\ \hline
\end{tabular}
}
\end{table}

\begin{table}[!h]
\renewcommand{\arraystretch}{1.3}
\centering
\caption{Stream Processing Workloads}
\label{stream_workloads}
\resizebox{\columnwidth}{!}{
\begin{tabular}{l|l|l}
\hline
\textbf{Workload} & \multicolumn{1}{c|}{\textbf{Description}} & \textbf{\begin{tabular}[c]{@{}l@{}}Input \\ data \\ stream\end{tabular}} \\ \hline
\begin{tabular}[c]{@{}l@{}}Streaming \\ Kmeans (Skm)\end{tabular} & \begin{tabular}[c]{@{}l@{}}uses streaming version of K-Means clustering algorithm\\  from Spark Mllib.\end{tabular} & \multirow{3}{*}{\begin{tabular}[c]{@{}l@{}}Numerical\\ Records\end{tabular}} \\ \cline{1-2}
\begin{tabular}[c]{@{}l@{}}Streaming \\ Linear \\ Regression \\ (Slir)\end{tabular} & \begin{tabular}[c]{@{}l@{}}uses streaming version of Linear Regression algorithm \\ from Spark Mllib.\end{tabular} &  \\ \cline{1-2}
\begin{tabular}[c]{@{}l@{}}Streaming \\ Logistic \\ Regression \\ (Slogr)\end{tabular} & \begin{tabular}[c]{@{}l@{}}uses streaming version of Logistic Regression algorithm \\ from Spark Mllib.\end{tabular} &  \\ \hline
\begin{tabular}[c]{@{}l@{}}Network \\ Word Count \\ (NWc)\end{tabular} & \begin{tabular}[c]{@{}l@{}}counts the number of words in  text,received from a\\  data server listening on a TCP socket every 2 sec and \\ print the counts on the screen. A data server is created \\ by running Netcat (a networking utility in Unix systems \\ for creating TCP/UDP connections)\end{tabular} & \multirow{5}{*}{\begin{tabular}[c]{@{}l@{}}Wikipe-\\ dia data\end{tabular}} \\ \cline{1-2}
\begin{tabular}[c]{@{}l@{}}Network \\ Grep (Gp)\end{tabular} & \begin{tabular}[c]{@{}l@{}}counts how many  lines,have the word the in them every\\ sec and prints the counts on the screen.\end{tabular} &  \\ \cline{1-2}
\begin{tabular}[c]{@{}l@{}}Windowed \\ Word Count \\ (WWc)\end{tabular} & \begin{tabular}[c]{@{}l@{}}generates every 10 seconds, word counts over the last \\ 30 sec of  data received on a TCP socket every 2 sec.\end{tabular} &  \\ \cline{1-2}
\begin{tabular}[c]{@{}l@{}}Stateful Word \\ Count (StWc)\end{tabular} & \begin{tabular}[c]{@{}l@{}}counts words cumulatively in text received from the net-\\ work every sec starting with initial value of word count.\end{tabular} &  \\ \cline{1-2}
\begin{tabular}[c]{@{}l@{}}Sql Word \\ Count (SqWc)\end{tabular} & \begin{tabular}[c]{@{}l@{}}Use DataFrames and SQL to count words in  text recei-\\ ved from the network every 2 sec.\end{tabular} &  \\ \hline
\begin{tabular}[c]{@{}l@{}}Click stream\\  Error Rate \\ Per Zip Code \\ (CErpz)\end{tabular} & \begin{tabular}[c]{@{}l@{}}returns the rate of error pages (a non 200 status) in each \\ zipcode over the last 30 sec. A page view generator gen-\\ erates streaming events over the network to simulate  \\ page views per second on a website.\end{tabular} & \multirow{5}{*}{\begin{tabular}[c]{@{}l@{}}Click \\ streams\end{tabular}} \\ \cline{1-2}
\begin{tabular}[c]{@{}l@{}}Click stream \\ Page Counts \\ (CPc)\end{tabular} & counts views per URL seen in each batch. &  \\ \cline{1-2}
\begin{tabular}[c]{@{}l@{}}Click stream\\ Active User \\ Count (CAuc)\end{tabular} & returns number of unique users in last 15 sec &  \\ \cline{1-2}
\begin{tabular}[c]{@{}l@{}}Click stream \\ Popular User \\ Seen (CPus)\end{tabular} & \begin{tabular}[c]{@{}l@{}}look for users in the existing dataset and print it\\ out if there is a match\end{tabular} &  \\ \cline{1-2}
\begin{tabular}[c]{@{}l@{}}Click stream \\ Sliding Page \\ Counts (CSpc)\end{tabular} & counts page views per URL in the last 10 sec &  \\ \hline
\begin{tabular}[c]{@{}l@{}}Twitter \\ Popular Tags\\ (TPt)\end{tabular} & \begin{tabular}[c]{@{}l@{}}calculates popular hashtags (topics) over sliding 10 and\\  60 sec windows from a Twitter stream.\end{tabular} & \multirow{3}{*}{\begin{tabular}[c]{@{}l@{}}Twitter \\ Stream\end{tabular}} \\ \cline{1-2}
\begin{tabular}[c]{@{}l@{}}Twitter \\ Count Min \\ Sketch (TCms)\end{tabular} & \begin{tabular}[c]{@{}l@{}}uses the Count-Min Sketch, from Twitter's Algebird \\ library, to compute windowed and global Top-K \\ estimates of user IDs occurring in a Twitter stream\end{tabular} &  \\ \cline{1-2}
\begin{tabular}[c]{@{}l@{}}Twitter \\ Hyper \\ Log Log (THll)\end{tabular} & \begin{tabular}[c]{@{}l@{}}uses HyperLogLog algorithm, from Twitter's Algebird\\  library, to compute a windowed and global estimate \\ of the unique user IDs occurring in a Twitter stream.\end{tabular} &  \\ \hline
\end{tabular}
}
\end{table}

\subsection{System Configuration}

Table~\ref{hardware} shows details about our test machine. Hyper-threading is only enabled during the evaluation of simultaneous multi-threading for Spark workloads. Otherwise Hyper-Threading and Turbo-boost are disabled through BIOS as per Intel Vtune guidelines to tune software on the Intel Xeon processor E5/E7 v2 family~\cite{e5_tuning}. With Hyper-Threading and Turbo-boost disabled, there are 24 cores in the system operating at the frequency of 2.7 GHz.

\begin{table}[!h]
\renewcommand{\arraystretch}{1.3}
\caption{Machine Details.}
\label{hardware}
\centering
\resizebox{\columnwidth}{!}{
\begin{tabular}{l|l|l}
\hline
\textbf{Component} & \multicolumn{2}{c}{\textbf{Details}} \\ \hline
Processor & \multicolumn{2}{l}{Intel Xeon E5-2697 V2, Ivy Bridge micro-architecture} \\ \hline
\multirow{6}{*}{} & Cores & 12 @ 2.7GHz (Turbo up 3.5GHz) \\ \cline{2-3} 
 & Threads & \begin{tabular}[c]{@{}l@{}}2 per Core (when Hyper-Threading \\ is enabled)\end{tabular} \\ \cline{2-3} 
 & Sockets & 2 \\ \cline{2-3} 
 & L1 Cache & \begin{tabular}[c]{@{}l@{}}32 KB for Instruction and \\ 32 KB for Data per Core\end{tabular} \\ \cline{2-3} 
 & L2 Cache & 256 KB per core \\ \cline{2-3} 
 & L3 Cache (LLC) & 30MB per Socket \\ \hline
Memory & \multicolumn{2}{l}{\begin{tabular}[c]{@{}l@{}}2 x 32GB, 4 DDR3 channels, Max BW 60GB/s\\ per Socket\end{tabular}} \\ \hline
OS & \multicolumn{2}{l}{Linux Kernel Version 2.6.32} \\ \hline
JVM & \multicolumn{2}{l}{Oracle Hotspot JDK 7u71} \\ \hline
Spark & \multicolumn{2}{l}{Version 1.5.0} \\ \hline
\end{tabular}
}
\end{table}

Table~\ref{parameters} also lists the parameters of JVM and Spark after tuning. For our experiments, we configure Spark in local mode in which driver and executor run inside a single JVM. We use HotSpot JDK version 7u71 configured in server mode (64 bit). The Hotspot JDK provides several parallel/concurrent GCs out of which we use  Parallel Scavenge (PS) and Parallel Mark Sweep for young and old generations respectively as recommended in~\cite{performance_spark_volume}. The heap size is chosen such that the memory consumed is within the system. The details on Spark internal parameters are available~\cite{spark_config}.
   
\begin{table}[]
\renewcommand{\arraystretch}{1.3}
\caption{Spark and JVM Parameters for Different Workloads.}
\label{parameters}
\centering
\resizebox{\columnwidth}{!}{
\begin{tabular}{l|clclc}
\hline
\multirow{2}{*}{\textbf{Parameters}} & \multicolumn{4}{c|}{\textbf{\begin{tabular}[c]{@{}c@{}}Batch \\ Processing\\ Workloads\end{tabular}}} & \multirow{2}{*}{\textbf{\begin{tabular}[c]{@{}c@{}}Stream \\ Processing \\ Workloads\end{tabular}}} \\ \cline{2-5}
 & \multicolumn{2}{c|}{\textbf{\begin{tabular}[c]{@{}c@{}}Spark-Core,\\ Spark-SQL\end{tabular}}} & \multicolumn{2}{c|}{\textbf{\begin{tabular}[c]{@{}c@{}}Spark Mllib, \\ Graph X\end{tabular}}} &  \\ \hline
spark.storage.memoryFraction & \multicolumn{2}{c|}{0.1} & \multicolumn{2}{c|}{0.6} & 0.4 \\ \hline
spark.shuffle.memoryFraction & \multicolumn{2}{c|}{0.7} & \multicolumn{2}{c|}{0.4} & 0.6 \\ \hline
spark.shuffle.consolidateFiles & \multicolumn{5}{c}{true} \\ \hline
spark.shuffle.compress & \multicolumn{5}{c}{true} \\ \hline
spark.shuffle.spill & \multicolumn{5}{c}{true} \\ \hline
spark.shuffle.spill.compress & \multicolumn{5}{c}{true} \\ \hline
spark.rdd.compress & \multicolumn{5}{c}{true} \\ \hline
spark.broadcast.compress & \multicolumn{5}{c}{true} \\ \hline
Heap Size (GB) & \multicolumn{5}{c}{50} \\ \hline
Old Generation Garbage Collector & \multicolumn{5}{c}{PS Mark Sweep} \\ \hline
Young Generation Garbage Collector & \multicolumn{5}{c}{PS Scavenge} \\ \hline
\end{tabular}
}
\end{table}

\subsection{Measurement Tools and Techniques}

We configure Spark to collect GC logs which are then parsed to measure time (called real time in GC logs) spent in garbage collection. We rely on the log files generated by Spark to calculate the execution time of the benchmarks. We use Intel Vtune Amplifier~\cite{Vtune} to perform general micro-architecture exploration and to collect hardware performance counters. We use numactl~\cite{numactl} to control the process and memory allocation affinity to a particular socket. We use hwloc~\cite{hwloc} to get the CPU ID of hardware threads. We use msr-tools~\cite{msrtools} to read and write model specific registers (MSRs).    

All measurement data are the average of three measure runs; Before each run, the buffer cache is cleared to avoid variation in the execution time of benchmarks. Through concurrency analysis in Intel Vtune, we found that executor pool threads in Spark start taking CPU time after 10 seconds. Hence, hardware performance counter values are collected  after the ramp-up period of 10 seconds. For batch processing workloads, the measurements are taken for the entire run of the applications and for stream processing workloads, the measurements are taken for 180 seconds as the sliding interval and duration of windows  in  streaming workloads considered are much less than 180 seconds. 

We use top-down analysis method proposed by Yasin~\cite{Top_Down_Method_for_Counters} to study the micro-architectural performance of the workloads. Earlier studies on profiling on big data workloads shows the efficacy of this method in identifying the micro-architectural bottlenecks~\cite{deep_dive_data_analytics,performance_spark,profiling_warehouse}. Super-scalar processors can be conceptually divided into the "front-end" where instructions are fetched and decoded into constituent operations, and the "back-end" where the required computation is performed. A pipeline slot represents the hardware resources needed to process one micro-operation. The top-down method assumes that for each CPU core, there are four pipeline slots available per clock cycle. At issue point each pipeline slot is classified into one of four base categories: Front-end Bound, Back-end Bound, Bad Speculation and Retiring. If a micro-operation is issued in a given cycle, it would eventually either get retired or cancelled. Thus it can be attributed to either Retiring or Bad Speculation respectively. Pipeline slots that could not be filled with micro-operations due to problems in the front-end are attributed to Front-end Bound category whereas pipeline slot where no micro-operations are delivered due to a lack of required resources for accepting more micro-operations in the back-end of the pipeline are identified as Back-end Bound.   

The top-down method requires following the metrics described in Table~\ref{metrics_topdown}, whose definition are taken from Intel Vtune on-line help~\cite{Vtune}.

\begin{table}[]
\centering
\renewcommand{\arraystretch}{1.3}
\caption{Metrics for Top-Down Analysis of Workloads}
\label{metrics_topdown}
\resizebox{\columnwidth}{!}{
\begin{tabular}{l|l}
\hline
\textbf{Metrics} & \textbf{Description} \\ \hline
IPC & \begin{tabular}[c]{@{}l@{}}average number of retired instructions \\ per clock cycle\end{tabular} \\ \hline
DRAM Bound & \begin{tabular}[c]{@{}l@{}}how often CPU was stalled on the main\\  memory\end{tabular} \\ \hline
L1 Bound & \begin{tabular}[c]{@{}l@{}}how often machine was stalled without \\ missing the L1 data cache\end{tabular} \\ \hline
L2 Bound & \begin{tabular}[c]{@{}l@{}}how often machine was stalled on L2 \\ cache\end{tabular} \\ \hline
L3 Bound & \begin{tabular}[c]{@{}l@{}}how often CPU was stalled on L3 cache, \\ or contended with a sibling Core\end{tabular} \\ \hline
Store Bound & \begin{tabular}[c]{@{}l@{}}how often CPU was stalled on store \\ operations\end{tabular} \\ \hline
Front-End Bandwidth & \begin{tabular}[c]{@{}l@{}}fraction of slots during which CPU was \\ stalled due to front-end bandwidth issues\end{tabular} \\ \hline
Front-End Latency & \begin{tabular}[c]{@{}l@{}}fraction of slots during which CPU was \\ stalled due to front-end latency issues\end{tabular} \\ \hline
ICache Miss Impact & \begin{tabular}[c]{@{}l@{}}fraction of cycles spent on handling \\ instruction cache misses\end{tabular} \\ \hline
DTLB Overhead & \begin{tabular}[c]{@{}l@{}}fraction of cycles spent on handling \\ first-level data TLB load misses\end{tabular} \\ \hline
Cycles of 0 ports Utilized & \begin{tabular}[c]{@{}l@{}}the number of cycles during which\\ no port was utilized.\end{tabular} \\ \hline
\end{tabular}
}
\end{table}

\section{Evaluation}

\subsection{Does micro-architectural performance remain consistent across batch and stream processing data analytics?}

\begin{figure*}[]
\centering
\subfloat[IPC values of stream processing workloads lie in the same range as of batch processing workloads]{\includegraphics[scale=0.35]{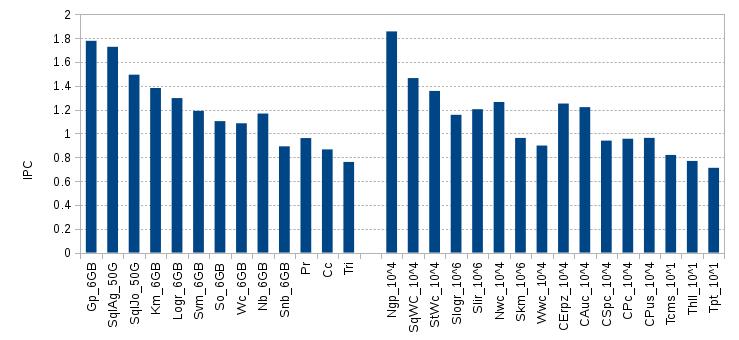}
\label{bs_ipc}}
\subfloat[Majority of stream processing workloads are back-end bound as that of batch processing workloads]{\includegraphics[scale=0.35]{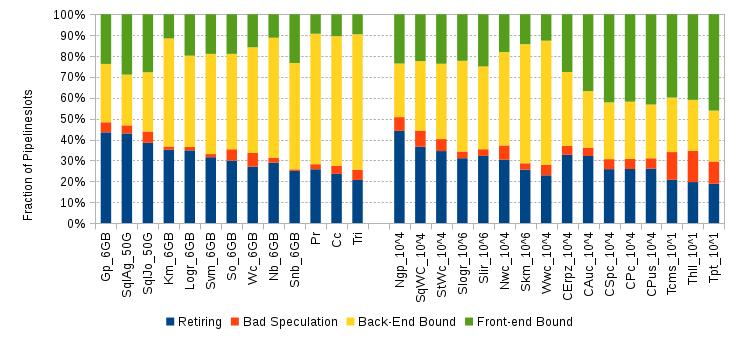}
\label{bs_topdown}}
\hfill
\subfloat[Stream processing workloads are also DRAM bound but their fraction of DRAM bound stalled cycles is lower than that of batch processing workloads]{\includegraphics[scale=0.35]{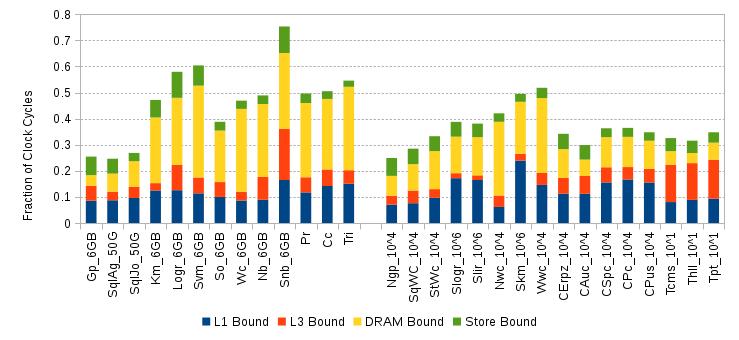}
\label{bs_memorybound}}
\subfloat[Memory bandwidth consumption of machine learning  based batch processing workloads is higher than other Spark workloads]{\includegraphics[scale=0.35]{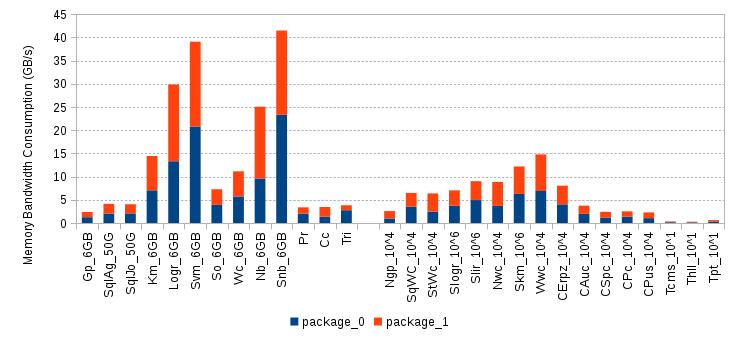}
\label{bs_bw}}
\hfill
\subfloat[Execution units starve both in batch in stream processing workloads]{\includegraphics[scale=0.35]{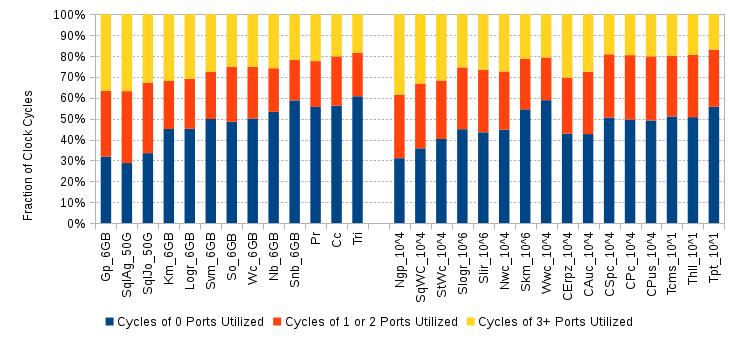}
\label{bs_cyclesports}}
\subfloat[ICache miss impact in majority of stream processing workloads is similar to batch processing workloads]{\includegraphics[scale=0.35]{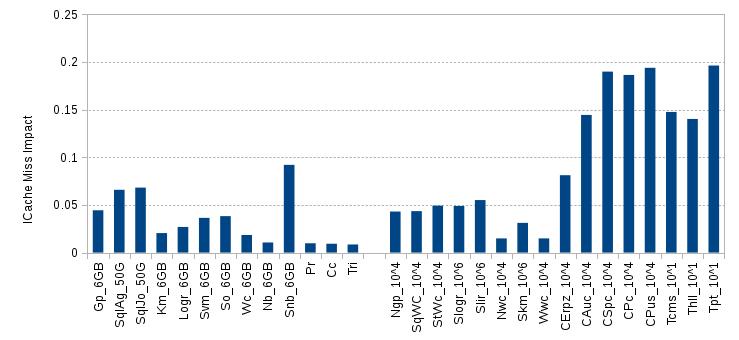}
\label{bs_icachemisses}}
\caption{Comparison of micro-architectural characteristics of batch and stream processing workloads}
\label{bs_compare}
\end{figure*}

As stream processing is micro-batch processing in Spark, we hypothesize batch processing and stream processing to exhibit same micro-architectural behaviour. Figure~\ref{bs_ipc} shows the IPC values of batch processing workloads range between 1.78 to 0.76, where as  IPC values of stream processing workloads also range between 1.85 to 0.71. The IPC values of word count (Wc) and grep (Gp) are very close to their stream processing equivalents, i.e. network word count (NWc) and network grep (NGp). Likewise the pipeline slots breakdown in Figure~\ref{bs_topdown} for the same workloads are quite similar. This implies that batch processing and stream processing will have same micro-architectural behaviour if the difference between two implementations is of micro-batching only. 

Sql Word Count(SqWc), which uses the Dataframes has better IPC than both word count (Wc) and network word count (NWc), which use RDDs. Aggregration (SqlAg) and Join (SqlAg) queries which also use DataFrame API have IPC values higher than most of the workloads using RDDs. One can see the similar pattern for retiring slots fraction in Figure~\ref{bs_topdown}. Sql Word Count (SqWc) exhibits 25.56\% less back-end bound slots than streaming network word count (NWc) because  sql word count (SqWc) shows 64\% less DRAM bound stalled cycles than network word count (NWc) and hence consumes 25.65\% less memory bandwidth than network word count (NWc). Moreover the execution units inside the core are less starved in sql word count as fraction of clock cycles during which no ports are utilized, is  5.23\% less than in network wordcount. RDDs use Java-objects based row representation, which have high space overhead  whereas DataFrames use new Unsafe Row format where rows are always 8-byte word aligned (size is multiple of 8 bytes) and equality comparison and hashing are performed on raw bytes without additional interpretation. This implies that Dataframes have the potential to improve the micro-architectural performance of Spark workloads.

The DAG of both windowed word count (Wwc) and  twitter popular tags (Tpt) consists of ``map'' and  ``reduceByKeyAndWindow'' transformations but the breakdown of pipeline slots in both workloads differ a lot. The back-end bound fraction in windowed word count (Wwc) is 2.44x larger and front-end bound fraction is 3.65x smaller than those in twitter popular tags (Tpt). The DRAM bound stalled cycles in windowed word count (Wwc) are 4.38x larger and L3 bound stalled cycles are 3.26x smaller than those in twitter popular tags (Tpt). Fraction of cycles during which 0 port is utilized, however differ only by 2.94\%. Icahce miss impact is 13.2x larger in twitter popular tags (Tpt) than in windowed word count (Wwc). The input data rate in windowed word count (Wwc) is 10,000 events/s whereas in twitter popular tags (Tpt), it is 10 events/s. Since the sampling interval is 2s, the working set of a windowing operation in windowed word count (Wwc) with 30s window length is 15 x 10,000 events where the working set of a windowing operation in twitter popular tags (Tpt) with 60s window length is 30 x 10 events. The working set in windowed word count (Wwc) is 500x larger than that in twitter popular tags (Tpt), The 30 MB last level cache is sufficient enough for the working set of Tpt but not for windowed word count (Wwc). That's why windowed word count (Wwc) also consumes 24x more bandwidth than twitter popular tags (Tpt). 

Click stream sliding page count (CSpc) also uses similar ``map'' and ``countByValueAndWindow'' transformations and the input data rate is also the same as in windowed word count (Wwc) but the back-end bound fraction and DRAM bound stalls are smaller in click stream sliding page count (CSpc) than in windowed word count (Wwc). Again the working set in Click stream sliding page count (CSpc) with 10s window length is 5 x 10,000 events which three times less than the working set in windowed word count (Wwc). 

CErpz and CAuc both use ``window'', ``map'' and ``groupbyKey'' transformations but the front-end bound fraction and icache miss imapct in CAuc is larger than in CErpz. However, back-end bound fraction, DRAM bound stalled cycles, memory bandwidth consumption are larger in CErpz than in CAuC. The retiring fraction is almost same in both workloads. The difference is again the working set. The working set in CErpz with window length of 30 seconds is 15 x 10,000 events which is 3x larger than in CAuc with window length of 10 seconds. This implies that with larger working sets, Icache miss impact can be reduced.

\subsection{How does data velocity affect micro-architectural performance of in-memory data analytics with Spark?}

\begin{figure*}[]
\centering
\subfloat[CPU utilization increases with data velocity]{\includegraphics[scale=0.40]{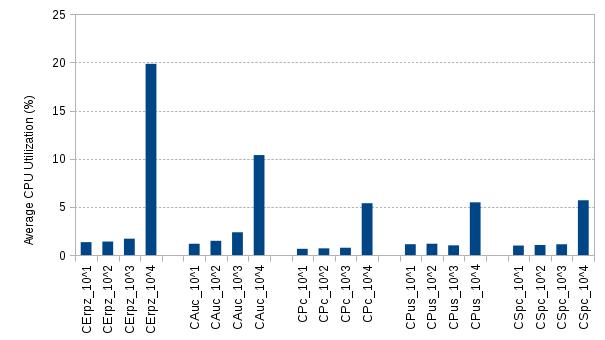}
\label{cpu_streaming_rate}}
\subfloat[Better IPC at higher data velocity]{\includegraphics[scale=0.40]{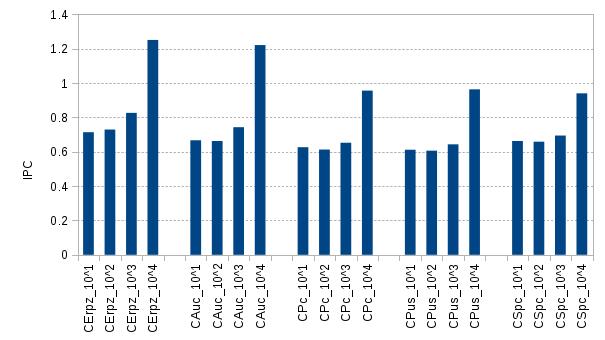}
\label{ipc_streaming_rate}}
\hfill
\subfloat[Front-end bound stalls decrease and fraction of retiring slots increases with data velocity]{\includegraphics[scale=0.40]{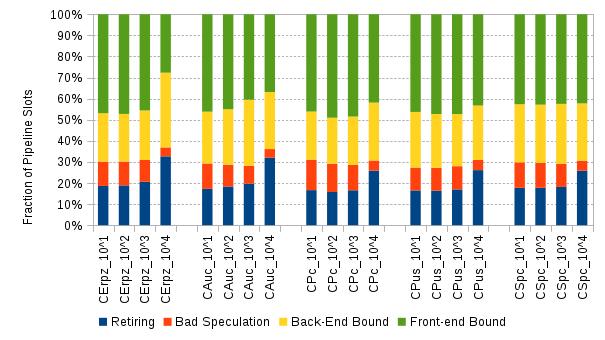}
\label{topdown_streaming_rate}}
\subfloat[Fraction of L1 Bound stalls increases, L3 Bound stalls decreases and DRAM bound stalls increases with data velocity]{\includegraphics[scale=0.40]{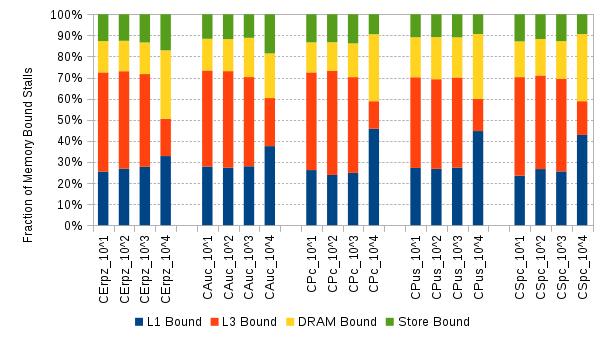}
\label{memorybound_streaming_rate}}
\hfill
\subfloat[Functional units inside exhibit better utilization at higher data velocity  ]{\includegraphics[scale=0.40]{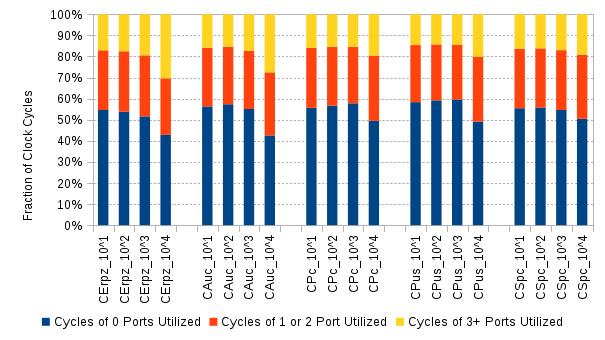}
\label{ports_streaming_rate}}
\subfloat[Memory bandwidth consumption increases with data velocity]{\includegraphics[scale=0.40]{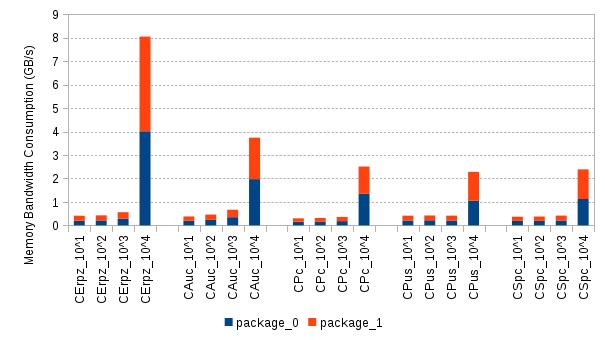}
\label{memorybw_streaming_rate}}
\caption{Impact of Data Velocity on Micro-architectural Performance of Spark Streaming Workloads}
\label{datavelocity}
\end{figure*}

In order to answer the question, we compare the micro-architectural characterization of stream processing workloads at input data rates of  10, 100, 1000 and 10,000 events per second. Figure~\ref{cpu_streaming_rate} shows that  CPU utilization increases only modestly up-to 1000 events/s after which it increases up-to 20\%. Like wise IPC in figure~\ref{ipc_streaming_rate} increases by 42\% in CSpc and 83\% in CAuc when input rate is increased from 10 to 10,000 events per second.

The pipeline slots breakdown in Figure~\ref{topdown_streaming_rate} shows that when the input data rates are increased from 10 to 10,000 events/s, fraction of pipeline slots being retired increases by 14.9\% in CAuc and 8.1\% in CSpc because in CAuc, the fraction of front-end bound slots and bad speculation slots decrease by 9.3\% and 8.1\% respectively and the back-end bound slots increase by only 2.5\%, where as in CSpc, the fraction of front-end bound slots and bad speculation slots decrease by 0.4\% and 7.4\% respectively and the back-end bound slots increase by only 0.4\%.       

The memory subsystem stalls break down in Figure~\ref{memorybound_streaming_rate} show that L1 bound stalls increase, L3 bound stalls decrease and DRAM bound stalls increase at high data input rate, e.g in CErpz, L3 bound stall and DRAM bound stalls remain roughly constant at 10, 100 and 1000 events/s because the working sets are still not large enough to create an impact but at 10,000 events/s, the working sets does not fit into the last level cache and thus DRAM bound stalls increase by approximately 20\% while the L3 bound stalls decrease by the same amount. This is also evident from Figure~\ref{memorybw_streaming_rate}, where the memory bandwidth consumption is constant at 10, 100 and 1000 events/s and then increases significantly at 10,000 events/s. Larger working sets translate into better utilization of functional units  as the number of clock cycles during which no ports are utilized decrease at higher input data rates. Hence input data rates should be high enough to provide working sets large enough to keep the execution units busy. 

\subsection{Does data locality on NUMA nodes improve the performance of in-memory data analytics with Spark?}

Ivy Bridge Server is a NUMA multi-socket system. Each socket has 2 on-chip memory controllers and a part of the main memory is directly connected to each socket. This layout offers high bandwidth and low access latency to the directly connected part of the main memory. The sockets are connected by two QPI (Quick Path Interconnect) links, thus a socket can access the main memory of other socket. However, a memory access from one socket to memory from another socket (remote memory access) incurs additional latency overhead due to transferring the data by cross-chip interconnect. By co-locating the computations with the data they access, the NUMA overhead can be avoided.     

To evaluate the impact of NUMA on Spark workloads, we run the benchmarks in two configurations: a) Local DRAM, where Spark process is bound to socket 0 and memory node 0, i.e. computations and data accesses are co-located, and b) Remote DRAM, where spark process is bound to socket 0 and memory node 1, i.e. all data accesses incur the additional latency. The input data size for the workloads is chosen as 6GB to ensure that memory working set sizes fit socket memory. Spark parameters for the two configurations are given in Table~\ref{numa}. Equation~\ref{eq:1} and~\ref{eq:2} in Appendix give the formulae for fraction of clock cycles, CPU stalled on local DRAM and remote DRAM respectively.

\begin{figure*}[!ht]
\centering
\subfloat[Performance degradation due to NUMA is 10\% on average across the workloads.]{\includegraphics[scale=0.40]{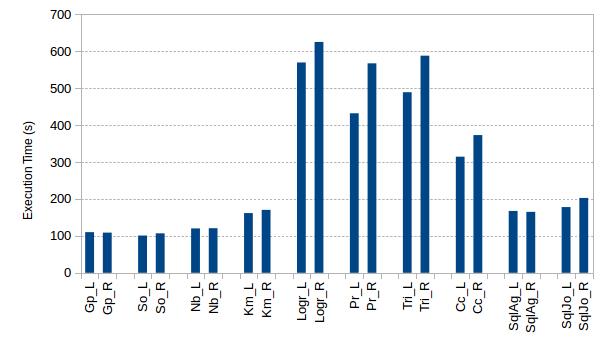}
\label{numa_penalty}}
\subfloat[Retiring decreases due to increased back-end bound in remote only mode.]{\includegraphics[scale=0.40]{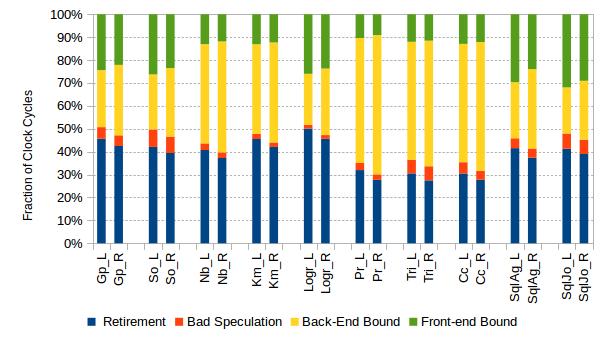}
\label{numa_topdown}}
\hfill
\subfloat[Stalled Cycles double in remote memory case]{\includegraphics[scale=0.40]{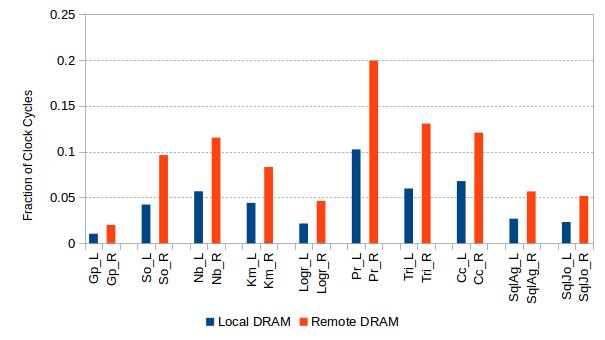}
\label{numa_local_dram}}
\subfloat[Memory Bandwidth consumption is well under the limits of QPI  bandwidth]{\includegraphics[scale=0.40]{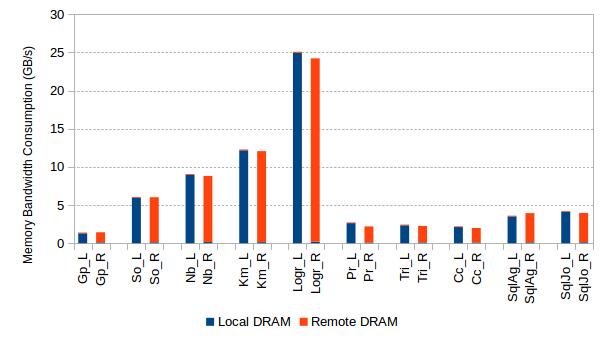}
\label{numa_bw}}
\caption{NUMA Characterization of Spark Benchmarks}
\label{numa_character}
\end{figure*}

Figure~\ref{numa_penalty} shows remote memory accesses can degrade the performance of Spark workloads by 10\% on average. This is because despite the stalled cycles on remote memory accesses double (see Figure~\ref{numa_local_dram}), retiring category degrades by only 8.7\%, Back-end bound stalls increases by 19.45\%, bad speculation decreases by 9.1\% and front-end bound stalls decreases by 9.58\% on average as shown in Figure~\ref{numa_topdown}. Furthermore the total cross-chip bandwidth of 32 GB/sec (peak bandwidth of 16 GB/s per QPI link) satisfies the memory bandwidth requirements of Spark workloads (see Figure~\ref{numa_bw}).

\begin{table}[!ht]
\renewcommand{\arraystretch}{1.3}
\centering
\caption{Machine and Spark Configuration for NUMA Evaluation}
\label{numa}
\resizebox{\columnwidth}{!}{
\begin{tabular}{l|l|c|c}
\multicolumn{2}{c|}{} & \textbf{Local DRAM} & \multicolumn{1}{l}{\textbf{Remote DRAM}} \\ \hline
\multirow{4}{*}{\textbf{Hardware}} & Socket ID & 0 & 0 \\ \cline{2-4} 
 & Memory Node ID & 0 & 1 \\ \cline{2-4} 
 & No. of cores & 12 & 12 \\ \cline{2-4} 
 & No. of threads & 12 & 12 \\ \hline
\multirow{3}{*}{\textbf{Spark}} & spark.driver.cores & 12 & 12 \\ \cline{2-4} 
 & spark.default.parallelism & 12 & 12 \\ \cline{2-4} 
 & spark.driver.memory (GB) & 24 & 24 \\ \hline
\end{tabular}
}
\end{table}

\subsection{Is simultaneous multi-threading effective for in-memory data analytics?}

Ivy Bridge Machine uses Simultaneous Multi-threading(SMT), which enables one processor core to run two software threads simultaneously to hide data access latencies. To evaluate the effectiveness of Hyper-Threading, we run Spark process in the three different configurations a) ST:2x1, the base-line single threaded configuration where Spark process is bound to two physical cores b) SMT:2x2, a simultaneous multi-threaded configuration where Spark process is allowed to use 2 physical cores and their corresponding hyper threads and c) ST:4x1, the upper-bound single threaded configuration where Spark process is allowed to use 4 physical cores. Spark parameters for the aforementioned configurations are given in Table~\ref{ht_config}. We also experimented with base-line configurations, ST:1x1, ST:3x3, ST:4x4, ST:5x5 and ST:6x6. In all experiments socket 0 and memory node 0 is used to avoid NUMA affects and the size of input data for the workloads is 6GB

\begin{table}[!ht]
\renewcommand{\arraystretch}{1.3}
\caption{Machine and Spark Configurations to evaluate Hyper Threading}
\label{ht_config}
\centering
\resizebox{\columnwidth}{!}{
\begin{tabular}{l|l|c|c|c}
\multicolumn{2}{c|}{} & \textbf{ST:2x1} & \multicolumn{1}{l|}{\textbf{SMT:2x2}} & \multicolumn{1}{l}{\textbf{ST:4x1}} \\ \hline
\multirow{4}{*}{\textbf{Hardware}} & No of  sockets & 1 & 1 & 1 \\ \cline{2-5} 
 & No of  memory nodes & 1 & 1 & 1 \\ \cline{2-5} 
 & No. of cores & 2 & 2 & 4 \\ \cline{2-5} 
 & No. of threads & 1 & 2 & 1 \\ \hline
\multirow{3}{*}{\textbf{Spark}} & spark.driver.cores & 2 & 4 & 4 \\ \cline{2-5} 
 & spark.default.parallelism & 2 & 4 & 4 \\ \cline{2-5} 
 & spark.driver.memory (GB) & 24 & 24 & 24 \\ \hline
\end{tabular}
}
\end{table}

\begin{figure*}[!ht]
\centering
\subfloat[Multi-core vs Hyper-Threading]{\includegraphics[scale=0.40]{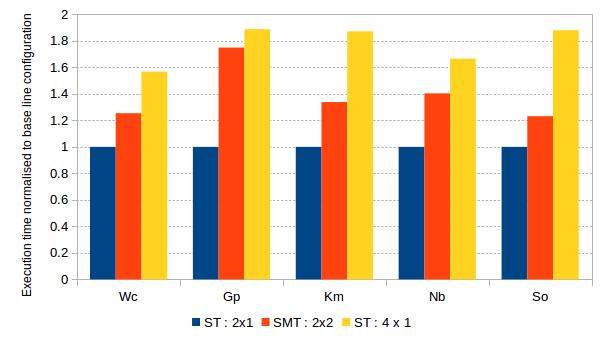}
\label{ht_speedup}}
\subfloat[HT Effectiveness is around 1]{\includegraphics[scale=0.40]{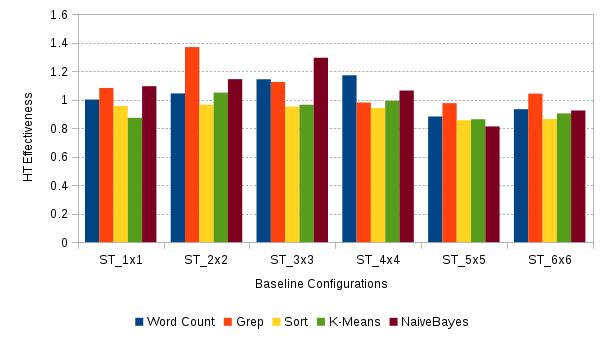}
\label{ht_cores}}
\hfill
\subfloat[Memory Bandwidth in multi-threaded case keeps up with that in multi-core case.]{\includegraphics[scale=0.40]{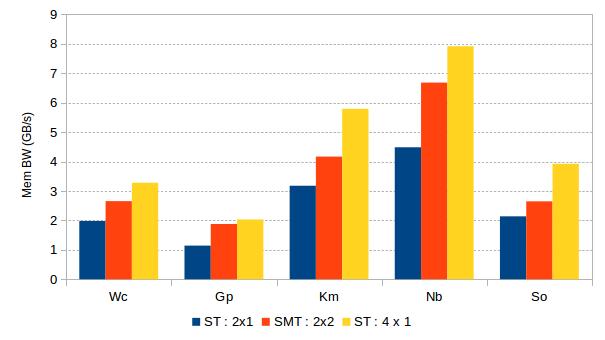}
\label{ht_bw}}
\subfloat[DRAM Bound decreases and L1 Bound increases]{\includegraphics[scale=0.40]{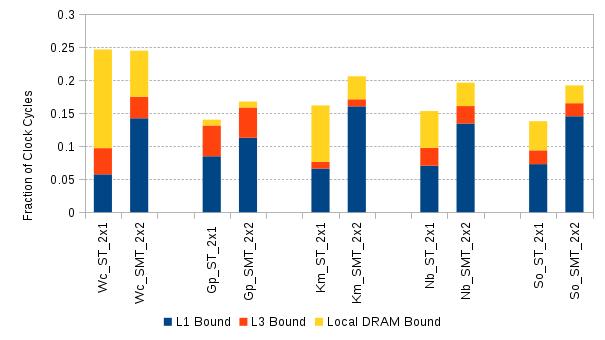}
\label{ht_memorybound}}
\caption{Hyper Threading is Effective}
\label{ht_effective}
\end{figure*}

Figure~\ref{ht_speedup} shows that SMT provides 39.5\% speedup on average across the workloads over baseline configuration, while the upper-bound configuration provided 77.45\% on average across the workloads. The memory bandwidth in SMT case also keeps up with multi-core case it is  20.54\% less than that of multi-core version on average across the workloads~\ref{ht_bw}. 
Figure~\ref{ht_cores} presents HT Effectiveness at different baseline configurations. HT Effectiveness of 1 is desirable as it implies 30\% performance improvement in Hyper-Threading case over the baseline single threaded configuration~\cite{ht_effectiveness}. Equation~\ref{eq:3} in Appendix gives the formula for HT effectiveness. One can see HT effectiveness remains close to 1 on average across the workloads till 4 cores after that it drops. This is because of poor multi-core scalability of Spark workloads as shown in~\cite{performance_spark}

For most of the workloads, DRAM bound is reduced to half whereas L1 Bound doubles when comparing the SMT case over baseline ST case in Figure~\ref{ht_memorybound} implying that Hyper-threading is effective in hiding the memory access latency for Spark workloads
 
\subsection{Are existing hardware prefetchers in modern scale-up servers effective for in-memory data analytics?}

Prefetching is a promising approach to hide memory access latency by predicting the future memory accesses  and fetching the corresponding memory blocks into the cache ahead of explicit accesses by the processor. Intel Ivy Bridge Server has two L1-D prefetchers and two L2 prefetchers.The description about prefetchers  is given  in Table~\ref{prefetchers}. This information is taken from Intel software forum~\cite{prefetchers}. 

\begin{table}[!ht]
\renewcommand{\arraystretch}{1.3}
\caption{Hardware Prefetchers Description}
\label{prefetchers}
\centering
\resizebox{\columnwidth}{!}{
\begin{tabular}{l|c|l}
\hline
\textbf{Prefetcher} & \textbf{\begin{tabular}[c]{@{}c@{}}Bit No. in \\ MSR\\ (0x1A4)\end{tabular}} & \multicolumn{1}{c}{\textbf{Description}} \\ \hline
\begin{tabular}[c]{@{}l@{}}L2 hardware \\ prefetcher\end{tabular} & 0 & \begin{tabular}[c]{@{}l@{}}Fetches additional lines of code \\ or data into the L2 cache\end{tabular} \\ \hline
\begin{tabular}[c]{@{}l@{}}L2 adjacent cache \\ line prefetcher\end{tabular} & 1 & \begin{tabular}[c]{@{}l@{}}Fetches the cache line thatcomprises\\  a cache line pair(128 bytes)\end{tabular} \\ \hline
DCU prefetcher & 2 & \begin{tabular}[c]{@{}l@{}}Fetches the next cache line into \\ L1-D cache\end{tabular} \\ \hline
DCU IP prefetcher & 3 & \begin{tabular}[c]{@{}l@{}}Uses sequential load history (based \\ on  Instruction Pointer of previous \\ loads) to determine whether to \\ prefetch additional lines\end{tabular} \\ \hline
\end{tabular}
}
\end{table}

To evaluate the effectiveness of L1-D prefetchers, we measure L1-D miss impact for the benchmarks at four configurations: a) all processor prefetchers are enabled, b) DCU prefetcher is disabled only, c) DCU IP prefetcher is disabled only and d) both L1-D prefetchers are disabled. To assess the effectiveness of L2 prefetchers, we measure L2 miss rate for the benchmarks at four configurations: a) all processor prefetchers are enabled, b) L2 hardware prefetcher is disabled only, c) L2 adjacent cache line prefetcher is disabled only and d) both L2 prefetchers are disabled. Equations~\ref{eq:4} and~\ref{eq:5} in the Appendix give formulae for L1-D miss impact and L2 hit impact.

\begin{figure*}[!ht]
\centering
\subfloat[L1-D DCU Prefetcher is ineffective]{\includegraphics[scale=0.40]{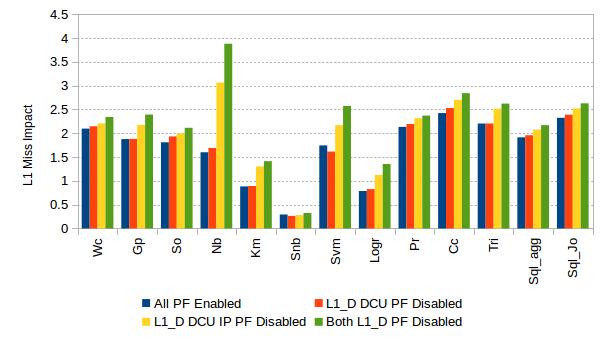}
\label{L1_prefetchers}}
\subfloat[Adjacent Cache Line L2 Prefecher is ineffective]{\includegraphics[scale=0.40]{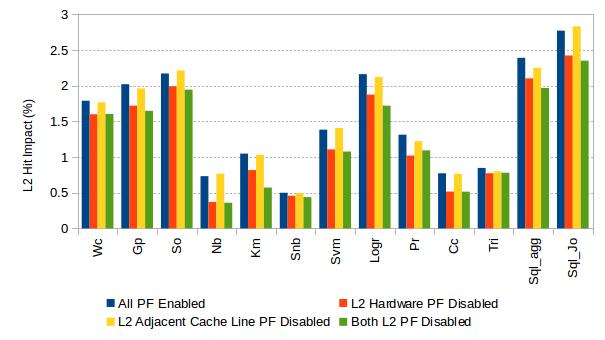}
\label{L2_prefetchers}}
\caption{Evaluation of Hardware Prefetchers}
\label{multiple_executors}
\end{figure*}

Figure~\ref{L1_prefetchers} shows that L1-D miss impact increases by only 3.17\% on average across the workloads when DCU prefetcher disabled, whereas the same metric increases by 34.13\% when DCU IP prefetcher is disabled in comparison with the case when all processor prefetchers are enabled. It implies that DCU prefetcher is ineffective.  

Figure~\ref{L2_prefetchers} shows that L2 hit impact decreases by 18\% on average across the workloads, when L2 adjacent cache line prefetcher disabled, whereas  disabling L2 adjacent line prefetcher decreases the L2 hit imapct by only 1.36\% on average across the workloads. This implies that L2 adjacent cache line prefetcher is ineffective.  .

\begin{figure}[!h]
\centering
\includegraphics[scale=0.40]{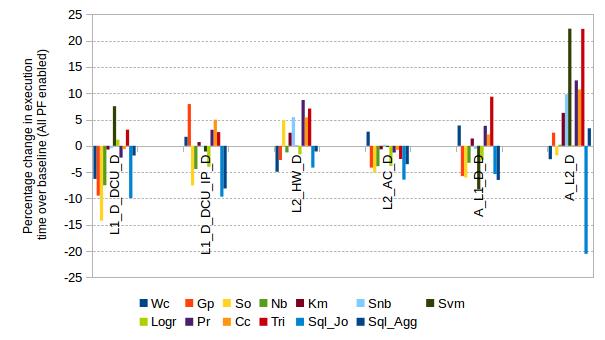}
\caption{Disabling L1-D next-line and L2 Adjacent Cache Line Prefetchers can reduce the execution of Spark jobs up-to 15\% and 5\% respectively }
\label{prefetchers_performance}
\end{figure}

By looking at the percentage change in execution time of Spark workloads over baseline configuration (all prefetchers are enabled), One can see that L1-D next-line and adjacent cache line L2 prefetchers have a negative impact on Spark workloads and disabling them improves the performance of Spark workloads up to 14.2\%  and 4.13\%. This shows that simple next-line hardware prefetchers in modern scale-up servers are ineffective for in-memory data analytics.

\subsection{Does in-memory data analytics with Spark experience loaded latencies (happens if bandwidth consumption is more than 80\% of sustained bandwidth)}

According to Jacob et al. in ~\cite{jacob2009memory}, the bandwidth vs latency response curve for a system has three regions. For the first 40\% of the sustained bandwidth, the latency response is nearly constant. The average memory latency equals idle latency in the system and the system performance is not limited by the memory bandwidth in the constant region. In between 40\% to 80\% of the sustained bandwidth, the average memory latency increases almost linearly due to contention overhead by numerous memory requests. The performance degradation of the system starts in this linear region. Between 80\% to 100\% of the sustained bandwidth, the memory latency can increase exponentially over the idle latency of DRAM system and the applications performance is limited by available memory bandwidth in this exponential region. Note that maximum sustained bandwidth is 65\% to 75\% of the theoretical maximum for server workloads.

Using the formula~\ref{eq:6}, taken from Intel's document~\cite{e5_tuning}, we calculate  that maximum theoretical bandwidth, per socket, for processor with DDR3-1866 and 4 channels is 59.7GB/s and the total system bandwidth is 119.4 GB/s. To find sustained maximum bandwidth, we compile the ompenmp version of STREAM~\cite{stream_benchmark} using Intel's icc compiler. The compiler flags used are given in the Appendix. On running the benchmark, we find maximum sustained bandwidth to be 92 GB/s.

\begin{figure}[!h]
\centering
\includegraphics[scale=0.40]{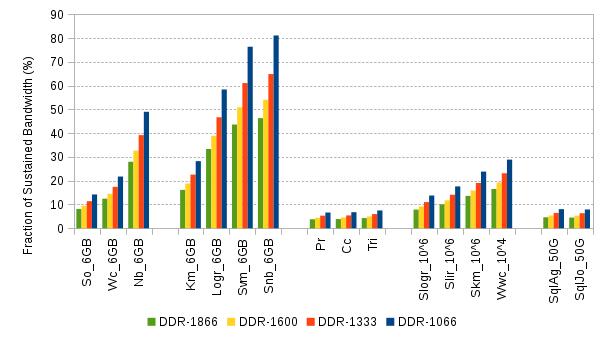}
\caption{Spark workloads don´t experience loaded latencies}
\label{fraction_sustained}
\end{figure}

\begin{figure}[!h]
\centering
\includegraphics[scale=0.27]{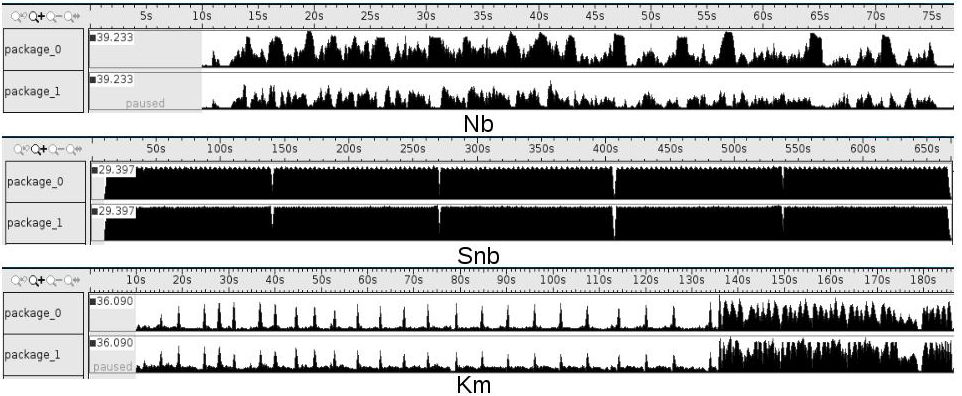}
\caption{Bandwidth Consumption over time}
\label{bw_time}
\end{figure}
    
Figure~\ref{fraction_sustained} shows the average bandwidth consumption as a fraction of sustained maximum bandwidth for different BIOS configurable data transfer rates of DDR3 memory. It can be seen that Spark workloads consume less than 40\% of sustained maximum bandwidth at 1866 data transfer rate and thus operate in the constant region. By lowering down the data transfer rates to 1066, majority of workloads from Spark core, all workloads from Spark SQL, Spark Streaming and Graph X still operate on the boundary of linear region where as workloads from Spark MLlib shift to the linear region and  mostly operate at the boundary of linear and exponential region. However at 1333, Spark MLlib workloads operate roughly in the middle of linear region. From the bandwidth consumption over time curves of the Km, Snb and Nb in Figure~\ref{bw_time},it can be seen that even when the peak bandwidth utilization goes into the exponential region, it lasts only for a short period of time and thus have negligible impact on the performance. 

It implies that Spark workloads do not experience loaded latencies and by lowering down the data transfer rate to 1333, performance is not affected. However, DRAM power consumption will be reduced as it is proportional to the frequency of DRAM.

\subsection{Are multiple small executors better than single large executor?}
With increase in the number of executors, the heap size of each executor's JVM is decreased. Heap size smaller than 32 GB enables ``CompressedOops'', that results in fewer garbage collection pauses. On the other-hand, multiple executors may need to communicate with each other and also with the driver. This leads to increase in the communication overhead. We study the trade-off between GC time and communication overhead for Spark applications. 

We deploy Spark in standalone mode on a single machine, i.e. master and worker daemons run on the same machine. We run applications with 1, 2, 4 and 6 executors. Beyond 6, we hit the operating system limit of maximum number of threads in the system. Table 1 lists downs the configuration details, e.g in 1E case, one Java Virtual Machine of 50 GB Heap size is launched and executor pool uses 24 threads, where as in 2E case 2 Java Virtual machines are launched, each with 25 GB of Heap space and 12 threads in the executor pool. In all configurations, the total number of cores and the total memory used by the applications are constant at 24 cores and 50GB respectively. 
 
\begin{table}[!h]
\centering
\renewcommand{\arraystretch}{1.3}
\caption{Multiple Executors Configuration}
\label{Multiple Executors Configuration}
\begin{tabular}{l|l|l|l|l}
\hline
\textbf{Configuration} & \textbf{1E} & \textbf{2E} & \textbf{4E} & \textbf{6E} \\ \hline
spark.executor.instances & 1 & 2 & 4 & 6 \\ \hline
spark.executor.memory (GB) & 50 & 25 & 12.5 & 8.33 \\ \hline
spark.executor.cores & 24 & 12 & 6 & 4 \\ \hline
spark.driver.cores & 1 & 1 & 1 & 1 \\ \hline
spark.driver.memory (GB) & 5 & 5 & 5 & 5 \\ \hline
\end{tabular}
\end{table}

\begin{figure}[!h]
\centering
\includegraphics[scale=0.40]{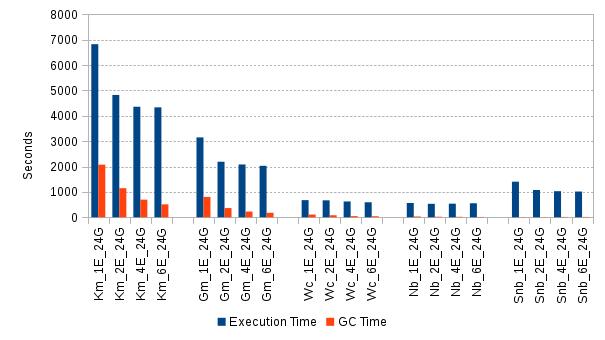}
\caption{Multiple small executors are better than single large executor due to reduction in GC time}
\label{multiple_executors}
\end{figure}

Figure~\ref{multiple_executors} data shows that 2 executors configuration are better than 1 executor configuration, e.g. for K-Means and Gaussian, 2E configuration provides 29.31\% and 30.43\% performance improvement over the baseline 1E configuration, however 6E configuration only increases the performance gain to 36.48\% and 35.47\% respectively. For the same workloads, GC time  in 6E case is 4.08x and and 4.60x less than 1E case. A small performance gain from 2E to 6E despite the reduction in GC time can be attributed to increased communication overhead among the executors and master

\section{Related Work}

Several studies characterize the behaviour of big data workloads and identify the mismatch between the processor and the big data applications~\cite{Clearing_the_clouds,DCBench,BigDataBench,understanding_in_memory_workloads,Characterising_and_subsetting_big_data_workloads, deep_dive_data_analytics, MMU_Performance_Scale_out_workloads}. However these studies lack in identifying the limitations of modern scale-up servers for Spark based data analytics. Ferdman et al.~\cite{Clearing_the_clouds} show that scale-out workloads suffer from high instruction-cache miss rates. Large LLC does not improve performance and off-chip bandwidth requirements of scale-out workloads are low. Zheng et al.~\cite{OS_behaviour_scale_out_workloads} infer that stalls due to kernel instruction execution greatly influence the front end efficiency. However, data analysis workloads have higher IPC than scale-out workloads~\cite{DCBench}. They also suffer from notable from end stalls but L2 and L3 caches are effective for them. Wang et al.~\cite{BigDataBench} conclude the same about L3 caches and L1 I Cache miss rates despite using larger data sets. Deep dive analysis~\cite{deep_dive_data_analytics} reveal that big data analysis workload is bound on memory latency but the conclusion can not be generalised. None of the above mentioned works consider frameworks that enable in-memory computing of data analysis workloads.

Tang et al.~\cite{google_numa} have shown that NUMA has significant impact on Gmail backend and web.search frontend.  Researchers at IBM's Spark technology center~\cite{Spark_numa_ibm} has only explored the thread affinity, bind only JVMs to sockets but does not limit the cross socket accesses. Beamer et al.~\cite{locality_graph} have shown NUMA has moderate performance penalty and SMT has limited potential for graph analytics running on Ivy bridge server. We show that exploiting the data locality on the modern servers will not yield significant performance gain for Spark and give micro-architectural reasons why this is so.

Kanev et al.~\cite{profiling_warehouse} have argued in favour of SMT after profiling live data center jobs on 20,000 google machines. While SMT has been shown to be effective for Hadoop workloads~\cite{deep_dive_data_analytics}, the same conclusion could not be translated about Spark workloads as previous work shows that as memory access characteristics of Spark and Hadoop differ~\cite{Characterising_and_subsetting_big_data_workloads} and software stacks have significant impact on the micro-architecture behaviour of big data workloads~\cite{Characterising_and_subsetting_big_data_workloads}. By reaching the same conclusion for Spark, we consolidate the general understanding of effectiveness of SMT for Big Data workloads

The general understanding about current Intel prefetchers is that they have either neutral or positive impact on SPEC benchmarks and Cloudsuite~\cite{Clearing_the_clouds}. We show for the first time the they have negative impact on the performance of in-memory data analytics with Spark.

\section{Conclusion}

We have reported a deep dive analysis of in-memory data analytics with Spark on a large scale-up server.

The key insights we have found are as follows:

\begin{itemize}
\item Batch processing and stream processing has same micro-architectural behaviour in Spark if the difference between two implementations is of micro-batching only. 
\item Spark workloads using DataFrames have improved instruction retirement over workloads using RDDs. 
\item If the input data rates are small, stream processing workloads are front-end bound. However, the front end bound stalls are reduced at larger input data rates and instruction retirement is improved. 
\item Exploiting data locality on NUMA nodes can only reduce the job completion time by 10\% on average as it reduces the back-end bound stalls by 19\%, which improves the instruction retirement only by 9\%. 
\item Hyper-Threading is effective to reduce DRAM bound stalls by 50\%, HT effectiveness is 1. 
\item Disabling next-line L1-D and Adjacent Cache line L2 prefetchers can improve the performance by up-to 14\% and 4\% respectively. 
\item Spark workloads does not experience loaded latencies and it is better to lower down the DDR3 speed from 1866 to 1333. 
\item Multiple small executors can provide up-to 36\% speedup over single large executor.
\end{itemize}

Firstly, we recommend Spark users to prefer DataFrames over RDDs while developing Spark applications and input data rates should be large enough for real time streaming analytics to improve the instruction retirement. Secondly, We advise to use executors with memory size less than or equal to 32GB and restrict each executor to use NUMA local memory. Thirdly we recommend to enable hyper-threading, disable next-line L1-D and adjacent cache line L2 prefetchers and lower the DDR3 speed to 1333. 

We also envision processors with 6 hyper-threaded cores without L1-D next line and adjacent cache line L2 prefetchers. The die area saved can be used to increase the LLC capacity. and the use of high bandwidth memories like Hybrid memory cubes is not justified for  in-memory data analytics with Spark.

\section{Appendix}
Here we give the formulas for metrics used in the evaluation of NUMA, SMT and hardware prefetchers in our study. 

Equation~\ref{eq:1} gives the formula for $Local DRAM Bound$, which tells how often the CPU was stalled on local memory node. It is calculated by multiplying the number of retired load micro-operations, which data sources missed LLC but serviced from local dram with the corresponding latency cycles and then dividing by the total number of  clock cycles when the cores are not in halted state. 

\begin{equation}
\label{eq:1}
\begin{split}
&Local\ DRAM\ Bound=(130*MEM\_LOAD\_UOPS\_LLC\\
&\_MISS\_RETIRED.LOCAL\_DRAM)/\\
&CPU\_CLK\_UNHALTED.THREAD
\end{split}    
\end{equation}

Equation~\ref{eq:2} gives the formula for $Remote DRAM Bound$, which tells how often the CPU was stalled on remote memory node. It is calculated by multiplying the number of retired load micro-operations, which data sources missed LLC but serviced from remote dram with the corresponding latency cycles and then dividing by the total number of  clock cycles when the cores are not in halted state. 

\begin{equation}
\label{eq:2}
\begin{split}
&Remote\ DRAM\ Bound=(310*MEM\_LOAD\_UOPS\_\\
&LLC\_MISS\_RETIRED.REMOTE\_DRAM)/\\
&CPU\_CLK\_UNHALTED.THREAD
\end{split}    
\end{equation}

Equation~\ref{eq:3} gives the formula for $HT\ Effectiveness$, which is taken from Intel's on-line forum~\cite{ht_effectiveness}. $HT\ Scaling_{obs}$ is the speedup observed in simultaneous multi-threaded case over the baseline single-threaded case, whereas $DP\ Scaling_{obs}$ speedup observed in the upper-bound single-threaded case. 

\begin{equation}
\label{eq:3}
\begin{split}
&HT\ Effectiveness = HT\ Scaling_{obs}*(0.538 + 0.462/\\
&DP\ Scaling_{obs}))
\end{split}
\end{equation}

Equation~\ref{eq:4} gives the formula for L1 miss impact, which is obtained by multiplying the number of retired load micro-operations which data sources following L1 data-cache miss with the corresponding latency cycles and dividing product by the total number of  clock cycles when the cores are not in halted state.

\begin{equation}
\label{eq:4}
\begin{split}
&L1\ Miss\ Impact=(6 * MEM\_LOAD\_UOPS\_RETIRED\\
&.L1\_MISS\_PS)/CPU\_CLK\_UNHALTED.THREAD
\end{split}    
\end{equation}

Equation~\ref{eq:5} gives the formula for L2 hit impact, which is obtained by multiplying the number of retired load micro-operations with L2 cache hits as data sources, with the corresponding latency cycles and dividing product by the total number of  clock cycles when the cores are not in halted state.

\begin{equation}
\label{eq:5}
\begin{split}
&L2\ Hit\ Impact=(12 * MEM\_LOAD\_UOPS\_RETIRED\\
&.L2\_HIT\_PS)/CPU\_CLK\_UNHALTED.THREAD
\end{split}    
\end{equation}

\begin{equation}
\label{eq:6}
\begin{split}
&Maximum\ Theoretical\ Bandwidth\ per\ socket\ (GB/s)=\\
&(<MT/s> * 8\ Bytes/clock * <num\ channels>)/1000 
\end{split}    
\end{equation}

\begin{equation}
\label{eq:7}
\begin{split}
&icc\ -O3\ -openmp\ -DSTREAM\_ARRAY\_SIZE=\\
&64000000\ -opt-prefetch-distance=64,8\\
&-opt-streaming-cache-evict=0\\
&-opt-streaming-stores\ always\ stream.c\\
&-o stream\_omp.64M\_icc
\end{split}    
\end{equation}

\section*{Acknowledgments}
This work was supported by  Erasmus Mundus Joint Doctorate in Distributed Computing (EMJD-DC) program  funded by the Education, Audiovisual and Culture Executive Agency (EACEA) of the European Commission. It was also supported by projects, Severo Ochoa SEV2015-0493 and TIN2015-65316 from the Spanish Ministry of Science and Innovation. We thank Ananya Muddukrishna for their comments on the first draft of the paper. We also thank the anonymous reviewers for their constructive feedback

\bibliographystyle{IEEEtran}
\small{
\raggedright
\bibliography{IEEEabrv,paper_cluster_v1}

\begin{thebibliography}{10}
\providecommand{\url}[1]{#1}
\csname url@samestyle\endcsname
\providecommand{\newblock}{\relax}
\providecommand{\bibinfo}[2]{#2}
\providecommand{\BIBentrySTDinterwordspacing}{\spaceskip=0pt\relax}
\providecommand{\BIBentryALTinterwordstretchfactor}{4}
\providecommand{\BIBentryALTinterwordspacing}{\spaceskip=\fontdimen2\font plus
\BIBentryALTinterwordstretchfactor\fontdimen3\font minus
  \fontdimen4\font\relax}
\providecommand{\BIBforeignlanguage}[2]{{%
\expandafter\ifx\csname l@#1\endcsname\relax
\typeout{** WARNING: IEEEtran.bst: No hyphenation pattern has been}%
\typeout{** loaded for the language `#1'. Using the pattern for}%
\typeout{** the default language instead.}%
\else
\language=\csname l@#1\endcsname
\fi
#2}}
\providecommand{\BIBdecl}{\relax}
\BIBdecl

\bibitem{Scale_up_vs_Scale_out_for_Hadoop}
R.~Appuswamy, C.~Gkantsidis, D.~Narayanan, O.~Hodson, and A.~I.~T. Rowstron,
  ``Scale-up vs scale-out for hadoop: time to rethink?'' in \emph{{ACM}
  Symposium on Cloud Computing, {SOCC}}, 2013, p.~20.

\bibitem{Phoenix_Rebirth}
R.~M. Yoo, A.~Romano, and C.~Kozyrakis, ``Phoenix rebirth: Scalable mapreduce
  on a large-scale shared-memory system,'' in \emph{Proceedings of IEEE
  International Symposium on Workload Characterization (IISWC)}, 2009, pp.
  198--207.

\bibitem{Tiled_mr}
R.~Chen, H.~Chen, and B.~Zang, ``Tiled-mapreduce: Optimizing resource usages of
  data-parallel applications on multicore with tiling,'' in \emph{Proceedings
  of the 19th International Conference on Parallel Architectures and
  Compilation Techniques}, ser. PACT '10, 2010, pp. 523--534.

\bibitem{Polymer}
K.~Zhang, R.~Chen, and H.~Chen, ``Numa-aware graph-structured analytics,'' in
  \emph{Proceedings of the 20th ACM SIGPLAN Symposium on Principles and
  Practice of Parallel Programming}.\hskip 1em plus 0.5em minus 0.4em\relax
  ACM, 2015, pp. 183--193.

\bibitem{Spark}
M.~Zaharia, M.~Chowdhury, T.~Das, A.~Dave, J.~Ma, M.~McCauly, M.~J. Franklin,
  S.~Shenker, and I.~Stoica, ``Resilient distributed datasets: A fault-tolerant
  abstraction for in-memory cluster computing,'' in \emph{Presented as part of
  the 9th USENIX Symposium on Networked Systems Design and Implementation (NSDI
  12)}, San Jose, CA, 2012, pp. 15--28.

\bibitem{performance_spark}
A.~Javed~Awan, M.~Brorsson, V.~Vlassov, and E.~Ayguade, ``Performance
  characterization of in-memory data analytics on a modern cloud server,'' in
  \emph{Big Data and Cloud Computing (BDCloud), 2015 IEEE Fifth International
  Conference on}.\hskip 1em plus 0.5em minus 0.4em\relax IEEE, 2015, pp. 1--8.

\bibitem{performance_spark_volume}
A.~J. Awan, M.~Brorsson, V.~Vlassov, and E.~Ayguade, \emph{Big Data Benchmarks,
  Performance Optimization, and Emerging Hardware: 6th Workshop, BPOE 2015,
  Kohala, HI, USA, August 31 - September 4, 2015. Revised Selected
  Papers}.\hskip 1em plus 0.5em minus 0.4em\relax Springer International
  Publishing, 2016, ch. How Data Volume Affects Spark Based Data Analytics on a
  Scale-up Server, pp. 81--92.

\bibitem{DStreams}
M.~Zaharia, T.~Das, H.~Li, S.~Shenker, and I.~Stoica, ``Discretized streams: an
  efficient and fault-tolerant model for stream processing on large clusters,''
  in \emph{Proceedings of the 4th USENIX conference on Hot Topics in Cloud
  Ccomputing}.\hskip 1em plus 0.5em minus 0.4em\relax USENIX Association, 2012,
  pp. 10--10.

\bibitem{google_numa}
L.~Tang, J.~Mars, X.~Zhang, R.~Hagmann, R.~Hundt, and E.~Tune, ``Optimizing
  google's warehouse scale computers: The numa experience,'' in \emph{High
  Performance Computer Architecture (HPCA2013), 2013 IEEE 19th International
  Symposium on}.\hskip 1em plus 0.5em minus 0.4em\relax IEEE, 2013, pp.
  188--197.

\bibitem{profiling_warehouse}
S.~Kanev, J.~P. Darago, K.~Hazelwood, P.~Ranganathan, T.~Moseley, G.-Y. Wei,
  D.~Brooks, S.~Campanoni, K.~Brownell, T.~M. Jones \emph{et~al.}, ``Profiling
  a warehouse-scale computer,'' in \emph{Proceedings of the 42nd Annual
  International Symposium on Computer Architecture}.\hskip 1em plus 0.5em minus
  0.4em\relax ACM, 2015, pp. 158--169.

\bibitem{tungsten}
Project tungsten.
  \url{https://databricks.com/blog/2015/04/28/project-tungsten-bringing-spark-closer-to-bare-metal.html}.

\bibitem{BigDataBench}
L.~Wang, J.~Zhan, C.~Luo, Y.~Zhu, Q.~Yang, Y.~He, W.~Gao, Z.~Jia, Y.~Shi,
  S.~Zhang, C.~Zheng, G.~Lu, K.~Zhan, X.~Li, and B.~Qiu, ``Bigdatabench: {A}
  big data benchmark suite from internet services,'' in \emph{20th {IEEE}
  International Symposium on High Performance Computer Architecture, {HPCA}},
  2014, pp. 488--499.

\bibitem{HiBench}
S.~Huang, J.~Huang, J.~Dai, T.~Xie, and B.~Huang, ``The hibench benchmark
  suite: Characterization of the mapreduce-based data analysis,'' in \emph{Data
  Engineering Workshops (ICDEW), 2010 IEEE 26th International Conference on},
  2010, pp. 41--51.

\bibitem{streambench}
R.~Lu, G.~Wu, B.~Xie, and J.~Hu, ``Stream bench: Towards benchmarking modern
  distributed stream computing frameworks,'' in \emph{Utility and Cloud
  Computing (UCC), 2014 IEEE/ACM 7th International Conference on}.\hskip 1em
  plus 0.5em minus 0.4em\relax IEEE, 2014, pp. 69--78.

\bibitem{perera2015solution}
S.~Perera and S.~Suhothayan, ``Solution patterns for realtime streaming
  analytics,'' in \emph{Proceedings of the 9th ACM International Conference on
  Distributed Event-Based Systems}.\hskip 1em plus 0.5em minus 0.4em\relax ACM,
  2015, pp. 247--255.

\bibitem{mllib}
X.~Meng, J.~Bradley, B.~Yavuz, E.~Sparks, S.~Venkataraman, D.~Liu, J.~Freeman,
  D.~Tsai, M.~Amde, S.~Owen \emph{et~al.}, ``Mllib: Machine learning in apache
  spark,'' \emph{arXiv preprint arXiv:1505.06807}, 2015.

\bibitem{BDGS}
Z.~Ming, C.~Luo, W.~Gao, R.~Han, Q.~Yang, L.~Wang, and J.~Zhan, ``{BDGS}: A
  scalable big data generator suite in big data benchmarking,'' in
  \emph{Advancing Big Data Benchmarks}, ser. Lecture Notes in Computer Science,
  2014, vol. 8585, pp. 138--154.

\bibitem{e5_tuning}
{Using Intel VTune Amplifier XE to Tune Software on the Intel Xeon Processor
  E5/E7 v2 Family}.
  \url{https://software.intel.com/en-us/articles/using-intel-vtune-amplifier-xe-to-tune-software-on-the-intel-xeon-processor-e5e7-v2-family}.

\bibitem{spark_config}
Spark configuration.
  \url{https://spark.apache.org/docs/1.5.1/configuration.html}.

\bibitem{Vtune}
{Intel Vtune Amplifier XE} 2013.
  \url{http://software.intel.com/en-us/node/544393}.

\bibitem{numactl}
Numactl. \url{http://linux.die.net/man/8/numactl}.

\bibitem{hwloc}
F.~Broquedis, J.~Clet-Ortega, S.~Moreaud, N.~Furmento, B.~Goglin, G.~Mercier,
  S.~Thibault, and R.~Namyst, ``hwloc: A generic framework for managing
  hardware affinities in hpc applications,'' in \emph{Parallel, Distributed and
  Network-Based Processing (PDP), 2010 18th Euromicro International Conference
  on}.\hskip 1em plus 0.5em minus 0.4em\relax IEEE, 2010, pp. 180--186.

\bibitem{msrtools}
msr-tools. \url{https://01.org/msr-tools}.

\bibitem{Top_Down_Method_for_Counters}
A.~Yasin, ``A top-down method for performance analysis and counters
  architecture,'' in \emph{2014 {IEEE} International Symposium on Performance
  Analysis of Systems and Software, {ISPASS}}, 2014.

\bibitem{deep_dive_data_analytics}
A.~Yasin, Y.~Ben-Asher, and A.~Mendelson, ``Deep-dive analysis of the data
  analytics workload in cloudsuite,'' in \emph{Workload Characterization
  (IISWC), IEEE International Symposium on}, Oct 2014, pp. 202--211.

\bibitem{ht_effectiveness}
{HT Effectiveness}.
  \url{https://software.intel.com/en-us/articles/how-to-determine-the-effectiveness-of-hyper-threading-technology-with-an-application}.

\bibitem{prefetchers}
{Hardware Prefetcher Control on Intel Processors}.
  \url{https://software.intel.com/en-us/articles/disclosure-of-hw-prefetcher-control-on-some-intel-processors}.

\bibitem{jacob2009memory}
B.~Jacob, ``The memory system: you can't avoid it, you can't ignore it, you
  can't fake it,'' \emph{Synthesis Lectures on Computer Architecture}, vol.~4,
  no.~1, pp. 1--77, 2009.

\bibitem{stream_benchmark}
{STREAM}. \url{https://www.cs.virginia.edu/stream/}.

\bibitem{Clearing_the_clouds}
M.~Ferdman, A.~Adileh, O.~Kocberber, S.~Volos, M.~Alisafaee, D.~Jevdjic,
  C.~Kaynak, A.~D. Popescu, A.~Ailamaki, and B.~Falsafi, ``Clearing the clouds:
  A study of emerging scale-out workloads on modern hardware,'' in
  \emph{Proceedings of the Seventeenth International Conference on
  Architectural Support for Programming Languages and Operating Systems}, ser.
  ASPLOS XVII, 2012, pp. 37--48.

\bibitem{DCBench}
Z.~Jia, L.~Wang, J.~Zhan, L.~Zhang, and C.~Luo, ``Characterizing data analysis
  workloads in data centers,'' in \emph{Workload Characterization (IISWC), IEEE
  International Symposium on}, 2013, pp. 66--76.

\bibitem{understanding_in_memory_workloads}
T.~Jiang, Q.~Zhang, R.~Hou, L.~Chai, S.~A. McKee, Z.~Jia, and N.~Sun,
  ``Understanding the behavior of in-memory computing workloads,'' in
  \emph{Workload Characterization (IISWC), IEEE International Symposium on},
  2014, pp. 22--30.

\bibitem{Characterising_and_subsetting_big_data_workloads}
Z.~Jia, J.~Zhan, L.~Wang, R.~Han, S.~A. McKee, Q.~Yang, C.~Luo, and J.~Li,
  ``Characterizing and subsetting big data workloads,'' in \emph{Workload
  Characterization (IISWC), IEEE International Symposium on}, 2014, pp.
  191--201.

\bibitem{MMU_Performance_Scale_out_workloads}
V.~Karakostas, O.~S. Unsal, M.~Nemirovsky, A.~Cristal, and M.~Swift,
  ``Performance analysis of the memory management unit under scale-out
  workloads,'' in \emph{Workload Characterization (IISWC), IEEE International
  Symposium on}, Oct 2014, pp. 1--12.

\bibitem{OS_behaviour_scale_out_workloads}
C.~Zheng, J.~Zhan, Z.~Jia, and L.~Zhang, ``Characterizing {OS} behavior of
  scale-out data center workloads,'' in \emph{The Seventh Annual Workshop on
  the Interaction amongst Virtualization,Operating Systems and Computer
  Architecture(WIVOSCA2013) held in conjunction with The 40th International
  Symposium on Computer Architecture}, 2013.

\bibitem{Spark_numa_ibm}
Spark executors love numa process affinity.
  \url{http://www.spark.tc/spark-executors-love-numa-process-affinity/}.

\bibitem{locality_graph}
S.~Beamer, K.~Asanovic, and D.~Patterson, ``Locality exists in graph
  processing: Workload characterization on an ivy bridge server,'' in
  \emph{Workload Characterization (IISWC), 2015 IEEE International Symposium
  on}.\hskip 1em plus 0.5em minus 0.4em\relax IEEE, 2015, pp. 56--65.

\end{thebibliography}
}

\end{document}